\documentclass[%
 aip,
 rsi,%
 amsmath,amssymb,
preprint,%
]{revtex4-1}

\usepackage{graphicx}
\usepackage{dcolumn}
\usepackage{bm}
\usepackage{natbib}

\begin{document}

\title{Design of a scanning gate microscope in a cryogen-free dilution refrigerator}

\author{M. Pelliccione} 
\author{A. Sciambi}
\affiliation{Department of Applied Physics, Stanford University, Stanford, California 94305-4045, USA}
\affiliation{SIMES, SLAC National Accelerator Laboratory, 2575 Sand Hill Road, Menlo Park, California 94025, USA}
\author{J. Bartel}
\affiliation{SIMES, SLAC National Accelerator Laboratory, 2575 Sand Hill Road, Menlo Park, California 94025, USA}
\affiliation{Department of Physics, Stanford University, Stanford, California 94305-4045, USA}
\author{A. Keller}
\affiliation{Department of Physics, Stanford University, Stanford, California 94305-4045, USA}
\author{D. Goldhaber-Gordon}
\affiliation{SIMES, SLAC National Accelerator Laboratory, 2575 Sand Hill Road, Menlo Park, California 94025, USA}
\affiliation{Department of Physics, Stanford University, Stanford, California 94305-4045, USA}

\date{\today}

\begin{abstract}
We report on our design of a scanning gate microscope housed in a cryogen-free dilution refrigerator with a base temperature of 15 mK. The recent increase in efficiency of pulse tube cryocoolers has made cryogen-free systems popular in recent years. However, this new style of cryostat presents challenges for performing scanning probe measurements, mainly as a result of the vibrations introduced by the cryocooler. We demonstrate scanning with root-mean-square vibrations of 0.8 nm at 3 K and 2.1 nm at 15 mK in a 1 kHz bandwidth with our design.\end{abstract}

\pacs{07.79.Lh, 07.20.Mc}
\maketitle

\section{\label{sec:introduction}INTRODUCTION}
Scanning probe microscopy has been used as a very powerful tool to investigate mesoscopic systems at cryogenic temperatures\citep{Eriksson1996}, including electron hole puddles in graphene\citep{Martin2007}, and charging in quantum hall systems\citep{Tessmer1998} among others. In particular, scanning gate microscopy (SGM), which typically involves scanning a conductive AFM or STM tip over a sample, has shown the ability to probe branched electron flows through a quantum point contact\citep{Topinka2001,Jura2007} and to image islands of an Aharonov-Bohm interferometer\citep{Hackens2010}. Many SGM measurements are performed at liquid helium temperatures, but there has been a strong interest in performing SGM in a dilution refrigerator, which is desirable for improved energy resolution for spectroscopic measurements, as well as investigating physical effects that are absent at higher temperatures.

In recent years, a number of groups have developed scanning probe systems operating at dilution refrigerator temperatures, typically at or below 100 mK\citep{Huber2008,Gildemeister2007,Pelekhov1998,Moussy2001,Crook2002,Hedberg2010}. Most systems described in the literature are housed in traditional ``wet'' cryostats that utilize liquid helium (LHe) as a thermal bath. However, with the increasing cost and scarcity of liquid helium, many new cryostats are now ``dry'' systems in the sense that there is no liquid helium bath to provide cooling to 4.2 K. A LHe bath is instead replaced with a cryocooler, typically a pulse tube\citep{Wang2002,Wang1997}, which can provide conductive cooling to approximately 3 K. A traditional dilution unit based on the circulation of a $^{3}$He/$^{4}$He mixture is then used to reach temperatures on the order of 10 mK. These new systems are typically referred to as ``cryogen-free'' because of the absence of a LHe bath, although the systems do still rely on a $^{3}$He/$^{4}$He mixture similar to a traditional dilution system. Dry systems offer significant advantages over wet systems, including no He loss and the capacity for a large experimental volume at base temperature.

A significant disadvantage of a pulse tube based cryostat is the vibration introduced by the cryocooler\citep{Tomaru2005,Li2005,Ikushima2008,Caparrelli2006}. Although a pulse tube has no moving mechanical parts at low temperature like a Gifford-McMahon cryocooler\citep{Thirumaleshwar1986}, cooling is achieved by pulsing high pressure helium gas through the pulse tube at a frequency on the order of 1 Hz. Therefore, the nature of vibrations in the system is much different than those in LHe based system, where the vibrations from the thermal bath are a result of the boiling liquid cryogen. In this paper, we present our design of a scanning gate microscope in a ``cryogen-free'' dilution refrigerator. We built this microscope to perform measurements aimed at realizing local tunneling in high mobility 2D electron systems using a technique called ``virtual scanning tunneling microscopy'' (VSTM)\citep{Sciambi2010,Sciambi2011}.

\section{\label{sec:system}SYSTEM DESCRIPTION}
The SGM is housed inside a Leiden Cryogenics CF-650 dilution refrigerator, with a base temperature of 10 mK and 650 $\mu$W of cooling power at 120 mK with no experimental payload. Precooling to 3.3 K is provided by a Cryomech PT-415 pulse tube cryocooler driven by a CP1010 helium compressor, which itself provides 1.35 W of cooling power at 4.2 K with the remote motor option. A superconducting vector magnet from American Magnetics capable of applying 9 T perpendicular to the sample and 3 T parallel to the plane of the sample is integrated into the system and cooled by the pulse tube. The entire system is housed inside an acoustically and electrically shielded room built by ETS-Lindgren, which has 45 dB of RF attenuation at 1 GHz. All electrical penetrations into the room pass through pi filters from Spectrum Control (56-725-003 and 1289-004) which attenuate 70 dB at 1 GHz. The body of the scanner is composed of three coarse steppers (ANPx101 and ANPz101) and a piezo scanner (ANSxyz100) from Attocube Systems made of Grade 2 titanium.

The design of the system does not incorporate a top-loading probe, instead a sample is loaded into the system by disassembling the outer vacuum can, inner vacuum can, and radiation shields. The cooldown time for the system from room temperature to base temperature is approximately 36 hours, and it takes about 30 minutes to disassemble the system once warm to access the sample. The absence of a probe increases the turnaround time for changing samples, but simplifies the design considerably because it eliminates the space constraints set by the diameter of a probe. More importantly, it allows one to add significantly more mass to the scanning stage to improve vibration damping. The mixing chamber plate is 340 mm in diameter with a 125 mm clearance above the top of the magnet, which gives ample space for low temperature wiring and RF filtering.

Before presenting the specifics of the system design, it will be important to briefly describe the nature of vibrations introduced by the pulse tube. To achieve cooling, the helium gas pressure in the pulse tube must be pulsed at a particular frequency set by the dimensions of the pulse tube and the thermodynamics of the cooling cycle. For the PT-415 pulse tube used in this system, the pulse frequency is 1.40 Hz, and the gas pressure oscillates between 80 psi and 300 psi. The gas pulses are reasonably sharp in the sense that the vibration spectrum is reminiscent of the Fourier spectrum of a square wave, with harmonics measurable to well above 1 kHz. The integrated RMS vibrations up to 1 kHz on the mixing chamber plate are approximately 2 $\mu$m in the vertical $z$ direction, and 3 $\mu$m in the lateral directions $x$ and $y$. More details regarding the vibrations are presented in section \ref{vibrations}.

One straightforward way to reduce the vibrations from the pulse tube is to mechanically decouple it from the rest of the system. In our system, the pulse tube is bolted to a copper plate in the cryostat, which provides good thermal anchoring but also good mechanical anchoring. Replacing this joint with a mechanically floppy connection using copper braids or straps would help with vibrations, but would also compromise cooling power because of the increased thermal resistance of the joint. The PT-415 pulse tube provides only 1.35 W of cooling power at 4.2 K, which is significantly less than a LHe bath. Introducing a thermal impedance between the pulse tube and cryostat would effectively lower this value, which may be acceptable based on the cooling power requirements of the system. We have chosen not to isolate the pulse tube in this manner because of the cooling requirements of our vector magnet. 

Although we chose to not modify the thermal anchoring of the pulse tube to the cryostat, there are a few important points to highlight regarding the cold head motor. For the PT415 pulse tube, the cold head motor is a small stepper motor that turns a valve which cycles between the high and low pressure helium reservoirs supplied by the helium compressor. Vibrations from the rotation of this motor are important because the motor needs to be located close to the pulse tube to avoid a significant loss in cooling power. We use the remote motor option for the PT415 pulse tube, which allows the motor to be separated from the pulse tube by about 30 cm. However, this reduces the cooling power of the pulse tube from 1.5 W to 1.35 W at 4.2 K, and as such should not be moved much further away to avoid compromising more cooling power.

The mounting of the remote motor is also important because the flexible helium line between the motor and pulse tube expands and contracts as the helium pressure changes. For example, we found that mounting the remote motor on a platform separate from the fridge resulted in significant lateral vibrations, caused essentially by the motor ``rocking'' the fridge back and forth. As a result, we anchored the motor to one of the support legs of the fridge, but on a foam base that damps vibrations from the direct anchoring.

If the motor does not turn smoothly, the vibrations of the motor can add to the vibrations generated by the helium gas pulses. We found that the motor driver supplied by Cryomech (IM483) was far from ideal, generating vibrations in the stepper motor at 140 Hz and higher harmonics. The 140 Hz vibrations are a result of the design of the stepper motor; the stepper motor has 200 steps per revolution, and the motor turns at 0.7 Hz (one full rotation yields two gas pulses), and (200 steps per revolution) * (0.7 revolutions per second)  = 140 steps per second. The vibrations are a result of the nature of the stepper motor drive generated by the IM483. 

The cold head motor is a 4-wire bipolar stepper motor, and the ideal drive for a smooth rotation of the motor is a sinusoidal current through the windings. A typical microstepping driver, such as the IM483, discretizes the ideal sinusoidal current into microsteps approximating a sine wave, resulting in sharp steps of current through the motor windings and a somewhat ``jerky'' rotation. We also found this motor drive to be a significant source of radiated EMI due to a rapid change in current though the motor windings, which have an inductance of 15 mH per winding for this motor. The magnitude of the pickup will depend on the specific orientation and type of wiring installed, but we observed EMI in the 10 kHz to 1 MHz range with an amplitude above 10 mV on coaxial cabling installed in the fridge.

These problems can be improved dramatically by using a linear microstepping motor driver, which linearly ramps the current between microsteps instead of abruptly changing it. We modified the cold head motor to be driven by a custom LNX-G linear motor driver (Precision Motion Control, Inc.), which reduced the amplitude of 140 Hz vibrations at the mixing chamber plate by an order of magnitude. There is also no measurable difference in the noise on fridge wiring when comparing the noise with the cold head motor off and with the cold head motor on being driven by the linear driver.

\section{\label{sec:scanning}SGM DESIGN}
\begin{figure}[fig1]
\centering
\includegraphics[width=0.8\textwidth]{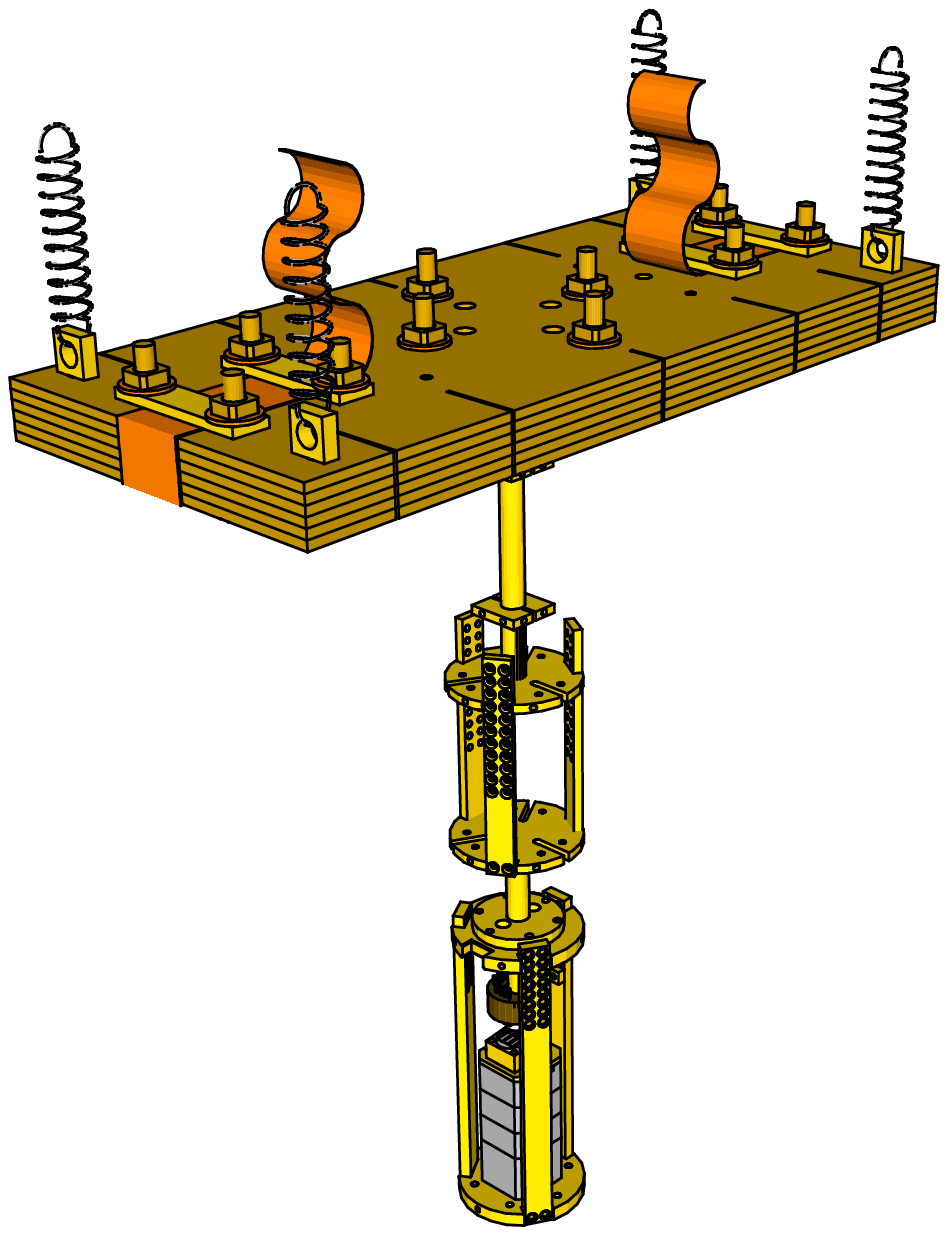}
\caption{Schematic drawing of the SGM suspended from the mixing chamber plate.}
\label{fig:1}
\end{figure}

\begin{figure}[fig2]
\centering
\includegraphics[width=0.8\textwidth]{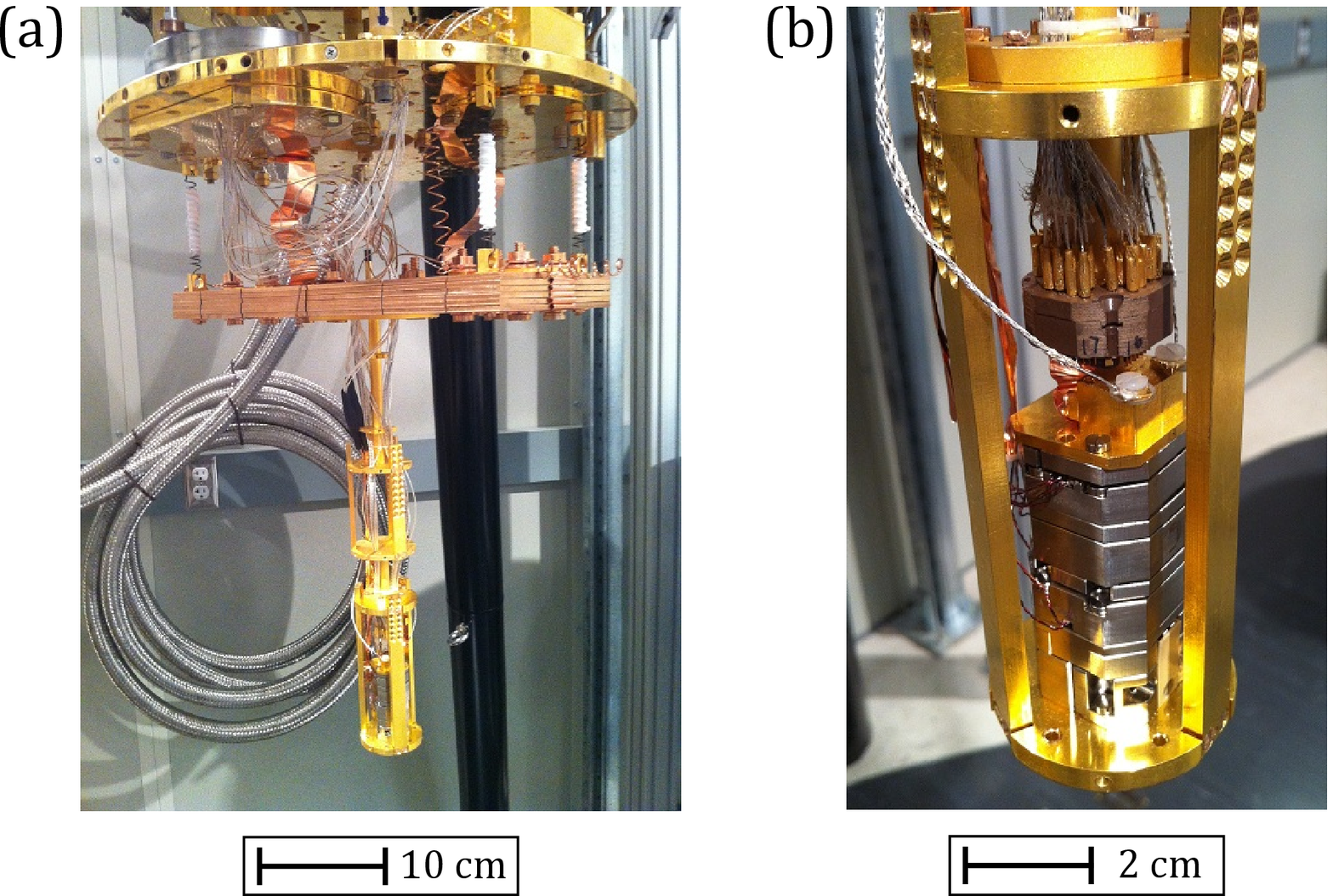}
\caption{(a) Picture of the SGM mounted on the mixing chamber plate of the CF650 dilution refrigerator. (b) Image of the scanning stage mounted at the bottom of the cold finger. The sample is seated facing down in the chip socket, and the tip rests on top of a stack of coarse steppers and a piezo scanner.}
\label{fig:2}
\end{figure}

The design of the SGM was chosen to attenuate vibrations from the pulse tube as much as possible. This is accomplished passively by implementing a low-pass spring stage in series with a mechanically stiff scanning stage, which will attenuate relative vibrations between the scanning tip and the sample. A diagram of the total ensemble is shown in figure \ref{fig:1} and consists of three main sections; a set of six 11''$\times$5''$\times$0.125'' phosphor bronze (PhBr) plates that hang via beryllium copper (BeCu) springs from the mixing chamber plate which we will refer to as the ``spring stage'', a gold plated 4N OFHC copper cold finger that extends to the center of the magnet, and the scanner mounted at the end of the cold finger. The cold finger is designed to have a variable height so the sample can be adjusted to rest in the field center of the magnet. A photograph of the SGM is shown in figure \ref{fig:2}(a), with a close-up image of the scanner in \ref{fig:2}(b).
\subsection{Spring Stage}
The choice of mass and spring constant for the spring stage is restricted by the dimensions of the system, particularly the 125 mm clearance between the bottom of the mixing chamber plate and the top of the magnet. The mass of the phosphor bronze plates is approximately 6 kg, a large mass helps reduce the amplitude of vibrations from the pulse tube. The resonant frequency of this stage is given by $f = (2\pi)^{-1}\sqrt{k/m} = (2\pi)^{-1}\sqrt{g/x_{eq}}$, where $x_{eq}$ is the equilibrium extension of the springs. The lowest resonant frequency achievable for the spring stage given the dimensions of the space is $f_{min} = (2\pi)^{-1}\sqrt{g/(125\mbox{ mm})} = 1.41\mbox{ Hz}$. Since the unstretched length of the springs and thickness of components is non-negligible, the maximum extension of the springs is less than 125 mm, and the actual resonant frequency of the stage in figure \ref{fig:2}(a) is 2.63 Hz at room temperature. Ideally, one would choose the resonant frequency of this stage to be considerably lower than the resonant frequency of the pulse tube for good attenuation. The relatively low pulse frequency of 1.40 Hz makes this difficult, and the spring stage does not act as an ideal low-pass filter for the pulses. It is important however that the resonance of the spring stage does not lie on a harmonic of the pulse frequency, or else vibrations could be amplified. 

The spring constant of the spring stage is approximately 1.75 N/mm at room temperature. This value may change when the stage is cooled due to thermal contraction of the BeCu and a change in the shear modulus. Given the shear modulus $G$, spring diameter $D$, wire radius $d$ and number of active coils $N_a$, the spring constant is given by\citep{Dimarogonas2000}
\[k = \frac{Gd^4}{8N_aD^3}.\]
For copper alloys, the shear modulus typically increases about 8\% from room temperature to 4 K\citep{Ledbetter1982}, and for BeCu the thermal contraction is about 0.3\% over the same temperature range\citep{Tuttle2010}. To first order, the spring constant at low temperature is then
\[k(\mbox{4 K}) \approx \frac{(1.08)(0.997)^4}{(0.997)^3}k(\mbox{300 K}) \approx (1.077)k(\mbox{300 K}).\]
This increase in spring constant will lead to an increase in the resonant frequency of the spring stage by a factor of $\sqrt{1.077}$ to 2.73 Hz, which is closer to the second harmonic of the pulse tube, but is still outside the bandwidth of the second harmonic, as will be shown in section \ref{vibrations}. Also, as can be seen in figure \ref{fig:2}(a), the BeCu springs are wrapped in Teflon tape. This provides damping to the springs without transmitting too much extra vibration. We found empirically that two layers of Teflon tape yielded the lowest vibrations at the scanner, extra layers of tape resulted in the springs being more rigid and increasing the vibrations.

\subsection{Thermal Anchoring}
Suspending the scanner via springs involves making a tradeoff between thermal anchoring and mechanical stiffness. Good thermal contact is achieved by rigidly anchoring the scanner to the mixing chamber plate, but this configuration results in tip-sample vibrations that are unacceptably large for our measurements. When tested in this configuration with our scanning cage, we observed over 200 nm peak-to-peak motion of the tip relative to the sample, with motion correlated with gas pulses from the pulse tube. This could perhaps be improved with a more rigid design for the scanning cage that houses the piezo scanner, but the stiffness is most likely limited by the relative rigidity of the Attocube coarse steppers and piezo scanner when compared to the copper scanning cage. 

\begin{figure}[fig3]
\centering
\includegraphics[width=0.8\textwidth]{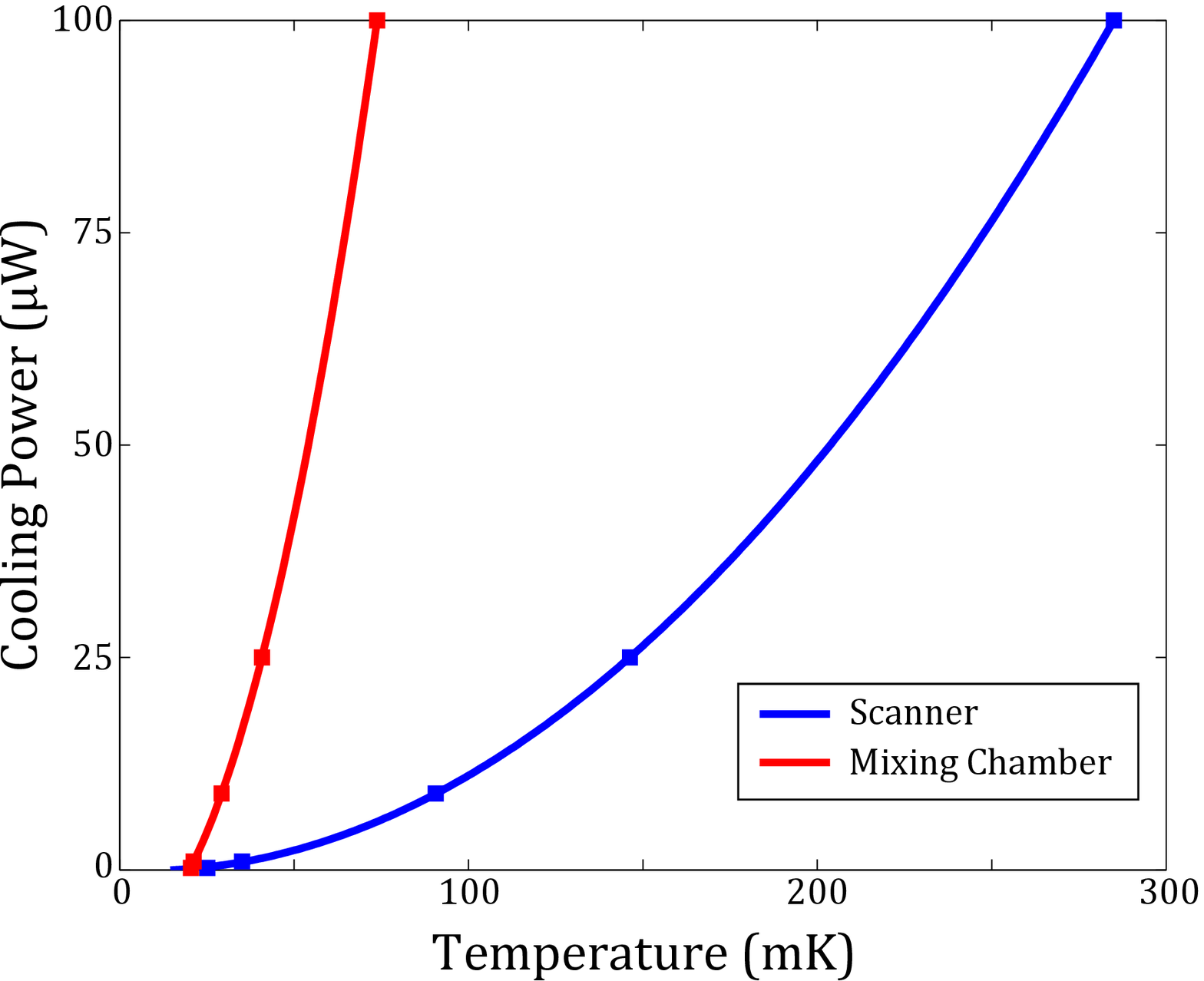}
\caption{Cooling power versus temperature for the mixing chamber and the scanner. The cooling power at the scanner is reduced to help mechanically decouple the scanner from the pulse tube. A fit to the predicted quadratic dependence of cooling power on temperature is plotted along with experimentally measured values.}
\label{fig:3}
\end{figure}

The design of this system has been biased towards achieving low vibration at the expense of thermal anchoring. The scanner is heat sunk to the mixing chamber plate with two 4N OFHC copper foils, each 5 mils thick. Although phosphor bronze has a lower thermal conductivity than copper, the spring stage has a large cross-section compared to the copper foil heat sinks, which are the dominant thermal impedance for cooling the SGM. To measure the effective cooling power at the scanner, heat is applied to the scanner and the temperature is measured both on the scanner and on the mixing chamber plate. The result of this measurement is shown in figure \ref{fig:3}.

The base temperature of the mixing chamber is 14.5 mK and the base temperature of the scanner is 15.4 mK. At $1\mbox{ }\mu$W of heat applied, the temperature of the scanner increases to 35 mK, and the mixing chamber temperature increases to 16.4 mK. The cooling power $\dot{Q}$ versus temperature $T$ can be extracted with a quadratic fit\citep{Pobell2007},
\[\dot{Q} - \dot{Q}_0 = a(T - T_0)^2,\]
where $\dot{Q}_0$ is the background heat load on the respective stage. Using the quadratic fitting, when the scanner is at 100 mK it has a cooling power of 11 $\mu$W, and the temperature of the mixing chamber plate is 31 mK. 

This is a considerable reduction in cooling power, but heat loads typically generated by the ANSxyz100 piezo scanner during large area scans are manageable at this scale. For example, a 30 minute long raster scan over an area of 3 $\mu$m $\times$ 3 $\mu$m  dissipates a power of about 1 $\mu$W due to losses in the scan piezos. The power $P$ dissipated in a piezoelectric element under the influence of a sinusoidal drive $V \sin(2\pi f t)$ is given by\citep{Thierauf2004}
\[P =2\pi f C V^2 \tan\delta,\]
where $C$ is the piezo capacitance and $\tan \delta$ is the dissipation factor due to dielectric loss in the piezoelectric material. Therefore, the heat load from the scanning piezos scales linearly with scan speed and quadratically with voltage or displacement. If a lower temperature is required during a scan, this can be achieved by either scanning more slowly, or over a smaller area. If a larger cooling power is required, the copper heat sinks could be replaced with 5N copper, or be subjected to an annealing procedure that improves the residual resistance ratio (RRR) at low temperatures.\citep{Risegari2004}

\subsection{Response in a Magnetic Field}
Integrating the spring supported SGM with a magnet presents a number of additional challenges. Since the spring stage is not well mechanically anchored to the rest of the system, extra care needs to be taken when operating in a large magnetic field to avoid forces that can lead to significant motion. This starts with choosing appropriate materials; not only avoiding ferromagnetic materials, but also materials with a large paramagnetic or diamagnetic response. This is especially true at base temperature where the magnetic susceptibility $\chi$ can follow a Curie-like $\chi\propto T^{-1}$ dependence. Materials that superconduct should also be avoided if possible because their mechanical response will be similar to a diamagnet with $\chi = -1$ until the magnetic field exceeds the critical field of the material.

If alloys are used, it is important to know the specific alloy composition since the magnetic response can vary significantly with small changes in elemental composition. One example that is important for our purposes are the various alloys of brass. Pure alloys of brass are alloys of copper and zinc only (e.g. C23000 and C26000), but most common alloys of brass include a non-negligible amount of lead (2.5-3.7\% for C36000)\citep{Davis2001}. Lead is added to improve the machinability of the brass, but unfortunately has a large effect on the magnetic response. Unleaded brass alloys have volume magnetic susceptibilities on the order of $\chi \sim 10^{-5}$ at 4 K, but leaded brass alloys can have $\chi \sim 10^{-2}$ at 4 K with only a few percent of lead added\citep{Fickett1992}. The leaded brass alloys also tend to have more relaxed constraints for the concentration of ferromagnetic elements including iron and nickel, although they are not an explicit component of the alloy. From our experience, it can be difficult to identify the alloy composition of a unknown brass part, as suppliers of components cannot always guarantee a specific alloy. However, by far the most common alloy chosen for machined brass products is C36000\citep{Kutz2002}. Unless the susceptibility is characterized with a low temperature SQUID, it is better to only use parts where the alloy is known. To this end, all anchoring hardware including screws, nuts, and bolts used on the SGM have been machined out of phosphor bronze (C51000), which has a susceptibility of $\chi = -5.6\times 10^{-6}$ at 4 K\citep{Fickett1992}.

As an example, consider the behavior of the spring stage if it was machined out of leaded brass instead of phosphor bronze. The force on a homogeneous material with volume magnetic susceptibility $\chi$ and volume $V$ in the presence of a field $\vec{B} = B(z)\hat{z}$ is
\begin{eqnarray*}
\vec{F} &=& \nabla \left(\vec{m}\cdot\vec{B} \right)\\
&=& \frac{\chi V}{\mu_0(1+\chi)}\nabla\left(\vec{B}\cdot\vec{B}\right)\\
&=& \frac{\chi V}{\mu_0(1+\chi)}\left(2B\frac{\partial B}{\partial z}\right)\hat{z}.
\end{eqnarray*}
Although the fringing fields of the magnet are not aligned along the $\hat{z}$ direction throughout the entire volume of the spring stage, we can make this assumption to simplify the calculation. Forces in the lateral directions should cancel due to symmetry since the SGM is centered along the axis of the solenoid. From simulations provided by American Magnetics, when the solenoid is energized to 9 T, the value of the $z$ component of the fringe field at the center of the spring stage is 0.6 T with a gradient of 0.1 T/cm in the $z$ direction. For phosphor bronze $\left(\mbox{C}51000, \chi = -5.6\times 10^{-6}\right)$ this gives a force $F = 0.036$ N, but for leaded brass $\left(\mbox{C}36000, \chi = 1.4\times 10^{-2}\right)$ this gives a force $F = 89$ N. Therefore, motion of the brass plate would be significant enough to cause a touch between the SGM and the fridge (51 mm) where it is not significant for the phosphor bronze (21 $\mu$m).

A similar calculation needs to be performed for the titanium in the coarse and fine positioners, which are machined of Grade 2 titanium\citep{AttocubeSystems} with $\chi = 1.5\times 10^{-4}$, and experience larger fields and field gradients than the spring  stage. Titanium will also superconduct at base temperature, although with a small critical field\citep{Falge1963} of $H_c\approx 5$ mT. For our design, forces on the titanium in these situations is negligible, yielding motion less than $100\mbox{ }\mu$m. In addition to titanium, any solder used in construction could superconduct and give unwanted motion. The only solder used in this design is silver solder in the cold finger, however the solder used (Silvaloy A-56T; 56\% Ag, 22\% Cu, 17\% Zn, 5\% Sn) has been shown not to superconduct\citep{Landau1972} down to 35 mK. Lead-tin solders have been avoided in the design, all joints in electrical wiring are either crimped or press-fit.

In addition to magnetic susceptibility, one must also consider the response of the SGM during magnetic field sweeps, and in the worst case in the event of a magnet quench. A rapid change in magnetic field will induce eddy currents in components with low electrical resistivity. Eddy currents in the presence of a magnetic field can result in   forces that move the SGM, and in the case of a magnet quench possibly cause damage. The force on a homogeneous material of circular cross-section in the presence of a time-varying magnetic field oriented normal to the circular cross-section is given by
\[\left|F\right| \approx \frac{\pi tr^4}{8\rho}\frac{dB}{dt}\left|\nabla B\right|,\]
where $r$ is the radius of the circular cross-section, $t$ is the thickness and $\rho$ is the electrical resistivity. To calculate the force on the spring stage in a magnet quench, assume it has a circular cross-section, and 1 second elapses from full field to zero field. Using the low temperature electrical resistivity of phosphor bronze $(\rho = 8.8\times 10^{-8}\mbox { }\Omega$m$)$\citep{Tuttle2010}, this gives a force  $F \approx 4$ N, which will move the SGM but is not enough to damage it. If the spring stage was instead made of 4N OFHC copper $(\rho = 1.6\times 10^{-10}$ $\Omega$m$)$\citep{Clark1970}, the force becomes $F \approx 2300$ N. 

This force is therefore a concern for any part made of OFHC copper, since it has a low electrical resistivity at cryogenic temperatures. However, the force due to eddy currents scales as $r^4$, and quickly becomes insignificant for parts with small dimensions. For example, the cold finger design includes a shaft with a diameter of 5/16" that rests inside the bore of the magnet. Even though it experiences a larger field and larger field gradient in this volume than the spring stage, the radius is small enough that the force it would experience in a magnet quench is not significant. The strong dependence on dimension is why this is typically not a concern for cold fingers mounted in probe based systems. The restriction on size imposed by the probe precludes the use of parts with lateral dimensions larger than a few inches. In those cases, the main motivation for reducing eddy currents is to reduce the heating associated with the dissipation of eddy currents in the materials. Eddy currents can also be reduced by cutting holes or slits into a part to impede eddy current flow. This is shown in figure \ref{fig:1} for the spring stage and the larger diameter copper parts in the cold finger. 

\subsection{Tuning Fork Cantilevers}
Operating the SGM in topographic mode is necessary to orient the location of the tip over the region of interest on a sample. This requires a scheme that allows for a measurement of the force (or force gradient) that the tip experiences. The most common method to accomplish this in AFM is a laser-based scheme, where a laser spot is reflected off the back of the cantilever, and the deflection of the laser is registered on a photodiode. There are two main complications with this method for cryogenic measurements, alignment and heating. If there is no optical access to the sample from room temperature, maintaining alignment of the laser and cantilever at cryogenic temperatures can be difficult, although not impossible. Residual heating from laser illumination is the larger problem at dilution temperatures, and for our measurements we would like to keep the heat load on the scanner under 1 $\mu$W. This restricts the laser intensity and the signal-to-noise ratio for the topographic measurement. In addition, mesoscopic samples, including those in the GaAs/AlGaAs material system, are typically light sensitive, and any unnecessary exposure to light at cryogenic temperatures should be avoided.

For these reasons, self-sensing cantilevers are of interest for cryogenic SGM since the use of a laser or other light source is not necessary. The most common self-sensing methods are piezoresistive cantilevers\citep{Harjee2010} and tuning fork cantilevers\citep{Giessibl2004,Giessibl1998,Giessibl2000,Rozhok2002}. Piezoresistive cantilevers are fabricated with a piezoelectric element in the cantilever that changes resistance as the cantilever is stressed. Although the fabrication of these types of cantilevers is more complicated than traditional silicon cantilevers, a measurement of the resistance change of the piezoresistor is relatively straightforward. A drawback of piezoresistve cantilevers for SGM in a dilution refrigerator is the heat dissipated by the piezoresistor. Typical sensitivities of piezoresistve cantilevers are a few ppm change in resistance per nanometer of deflection, and powers on the order of 1 mW are dissipated in the piezoresistor to achieve a suitable signal-to-noise ratio for imaging\citep{Harjee2010}. This power is manageable at 4 K or even in a $^3$He-based cryostat, but becomes problematic below 100 mK.

We have chosen to implement a tuning fork based cantilever scheme for topographic imaging. Tuning fork cantilevers were first introduced by Giessibl,\citep{Giessibl2004,Giessibl1998,Giessibl2000} and have sufficient force sensitivity to achieve atomic resolution under suitable conditions. Although we do not need atomic resolution imaging for our measurements, the very high force sensitivity of tuning forks is desirable. The interaction of the tip and sample is measured either with amplitude or frequency modulated techniques, with an ultimate deflection sensitivity on the order of 10 fm/$\sqrt{\mbox{Hz}}$ at 4 K\citep{Giessibl2000}. This is a result of the high quality factor $Q$ of the tuning fork, which can exceed 20000 in vacuum and at low temperature\citep{Giessibl2000}. Another consequence of the high quality factor at cryogenic temperatures is low power dissipation. We typically operate the tuning fork with powers much less than 1 nW, which is negligibly small for our measurements and has no measurable effect on the temperature of the scanner.

\begin{figure}[fig4]
\centering
\includegraphics[width=0.8\textwidth]{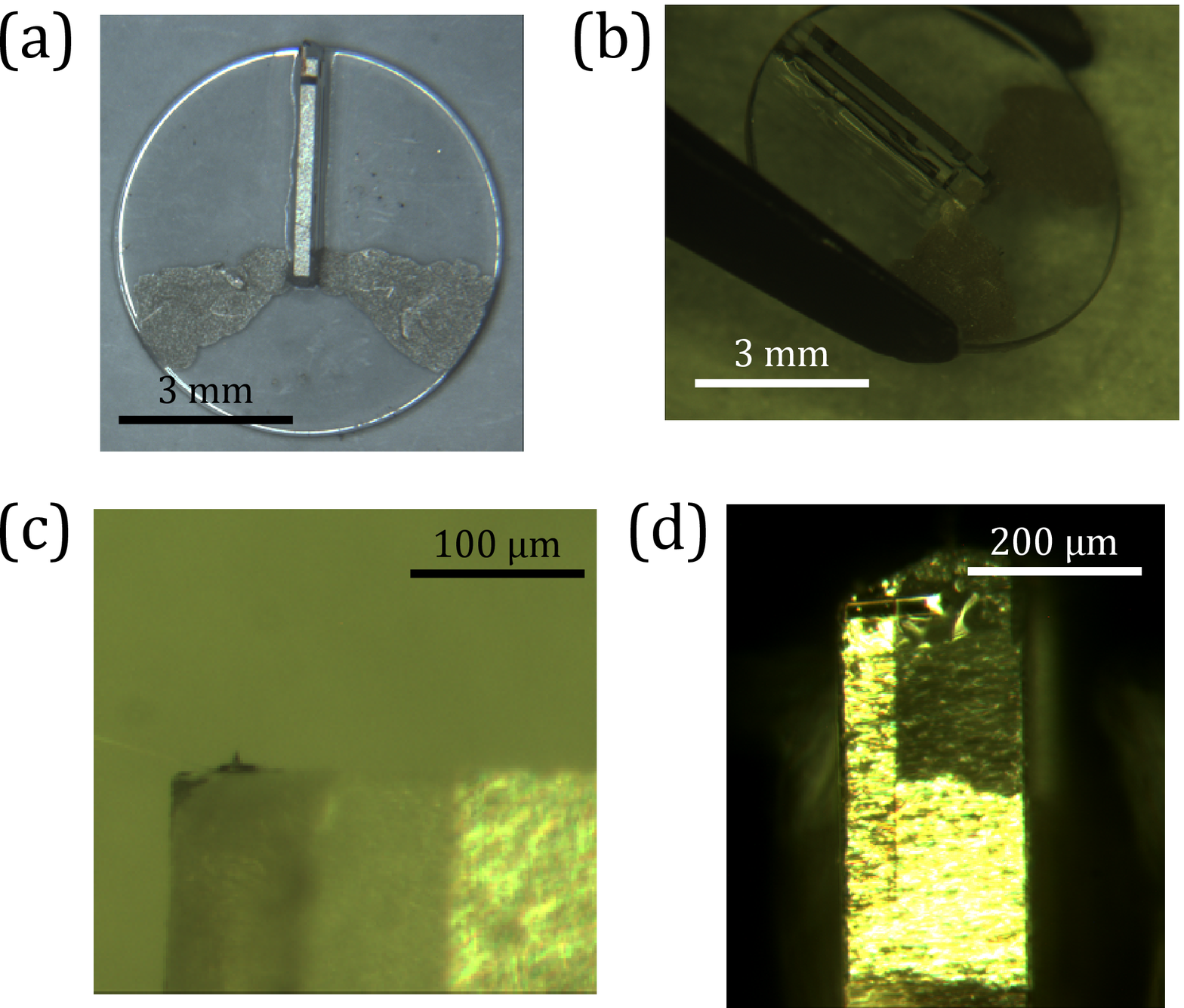}
\caption{Photographs of assembled tuning fork cantilevers. (a) Top view of the tuning fork anchored to a sapphire disk with silver contact pads. (b) Side view showing one prong anchored to the sapphire with adhesive. (c) Side view of a commercial AFM cantilever and tip attached to the end of the tuning fork. (d) Top view of the cantilever and tip epoxied to the end of the tuning fork with Ti/Au evaporated trace to make electrical contact to the tip.}
\label{fig:4}
\end{figure}
An image of a tuning fork cantilever with attached AFM tip is shown in figure \ref{fig:4}. The tuning forks are from IQD Frequency Products (LF A103C), and have a nominal resonant frequency of $2^{15} = 32768$ Hz. We fabricate cantilevers in the qPlus configuration, where one prong of the tuning fork is held fixed, and the other prong is free to vibrate. The tuning fork is mounted to a 7.5 mm diameter sapphire disk with cyanoacrylate glue (Krazy glue),\citep{Rozhok2002} where sapphire is chosen for its high thermal conductivity at low temperature for an electrical insulator. Electrical contact is made to the two leads of tuning fork with silver paint, which is used both to excite the mechanical resonance of the tuning fork and apply a voltage to the SGM tip when performing a scanning gate measurement. 

In the zoomed in views in figure \ref{fig:4}, the AFM tip and silicon cantilever can be seen mounted to the end of the upper prong of the tuning fork. The tip is attached using a fast-setting epoxy and a micromanipulator stage to controllably snap the silicon cantilever from the larger silicon piece it is manufactured with. The tips used are precoated in Pt with a 30 nm tip radius from MikroMasch (DPE14), and have a spring constant of 6 N/m. A shadow mask is then used to evaporate Ti/Au (20 nm/30 nm) on the end of the tuning fork to electrically connect the Pt-coated cantilever with one of the tuning fork leads. The result of this process is shown in figure \ref{fig:4}(d). The resonant frequency of the tuning fork after fabrication is slightly lower than its original resonance, typically 30 - 31 kHz, with a $Q$ around 750 in ambient conditions and 7500 at low temperature in vacuum.



\section{\label{sec:vibrations}VIBRATIONS}\label{vibrations}
The vibrations in the system are dominated by the pulse tube. We can characterize the vibrations using two different methods, one is simply measuring the deflection of the AFM tip while it is in tapping contact with the surface at a fixed position, which is sensitive to relative vibrations between the tip and sample. This measurement is most relevant for scanning since common mode vibrations of the scanner are not picked up.

To characterize the overall vibration floor of the system, it is more instructive to use geophones, which are simply a small magnet hanging from a spring coaxially inside a solenoid. When there is relative motion between the magnet and the solenoid, a voltage will be induced in the solenoid proportional to the velocity of the magnet for small displacements. We use geophones from GeoSpace LP, model GS-11D with a natural resonance frequency of 4.5 Hz. Geophones become less sensitive below their natural resonance because external vibrations result in a common mode motion of both the magnet and solenoid well below the resonant frequency. This yields a practical lower limit of 1 Hz for detectable vibration using this particular geophone. This model, however, is quite compact, and easily fits inside the system for measurements at room temperature and low temperature. 

\begin{figure}[fig5]
\centering
\includegraphics[width=0.8\textwidth]{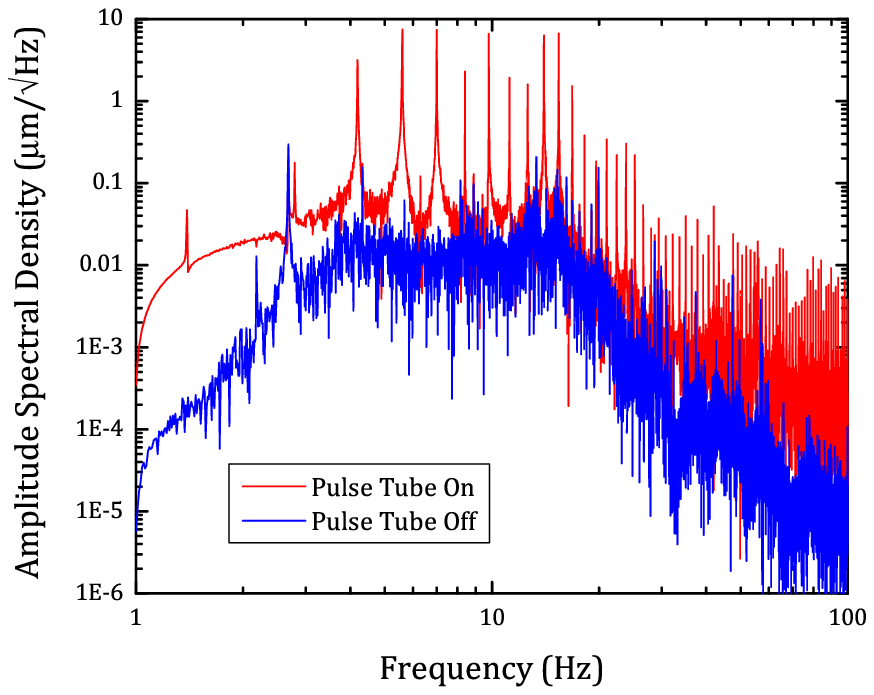}
\caption{Vibrations in the vertical direction on the mixing chamber plate as measured with a geophone at room temperature. Harmonics of the pulse frequency (1.40 Hz) are clearly visible in the spectrum when the pulse tube is on.}
\label{fig:5}
\end{figure}
Figure \ref{fig:5} shows the result of a vibration measurement in the vertical direction with a geophone on the mixing chamber plate at room temperature, comparing the vibrations with the pulse tube on and off. The effect of the pulse tube on vibrations is dramatic. We define the integrated RMS vibration $I$ between two frequencies $f_1$ and $f_2$ as
\[I(f_1,f_2) = \sqrt{\int_{f_1}^{f_2}\left[S(f)\right]^2df},\]
where $S(f)$ is the amplitude spectral density. Integrating the data in figure \ref{fig:5} from 1 Hz to 1 kHz gives an RMS vibration of 1.9 $\mu$m on the mixing chamber plate with the pulse tube on. The data in figure \ref{fig:5} is only plotted to 100 Hz to clearly show the full width at half maximum of the resonant peaks, which are approximately 25 mHz. Similar spectra were also taken at 3.3 K, the vibrations are qualitatively similar and the total RMS vibration changes only by a few percent. The harmonics of the pulse frequency (1.40 Hz) are measurable with the geophone up to 1 kHz, although the spectral weight of harmonics near 1 kHz is small, on the order of $10^{-5}$ $\mu$m/$\sqrt{\mbox{Hz}}$.

\begin{figure}[fig6]
\centering
\includegraphics[width=0.8\textwidth]{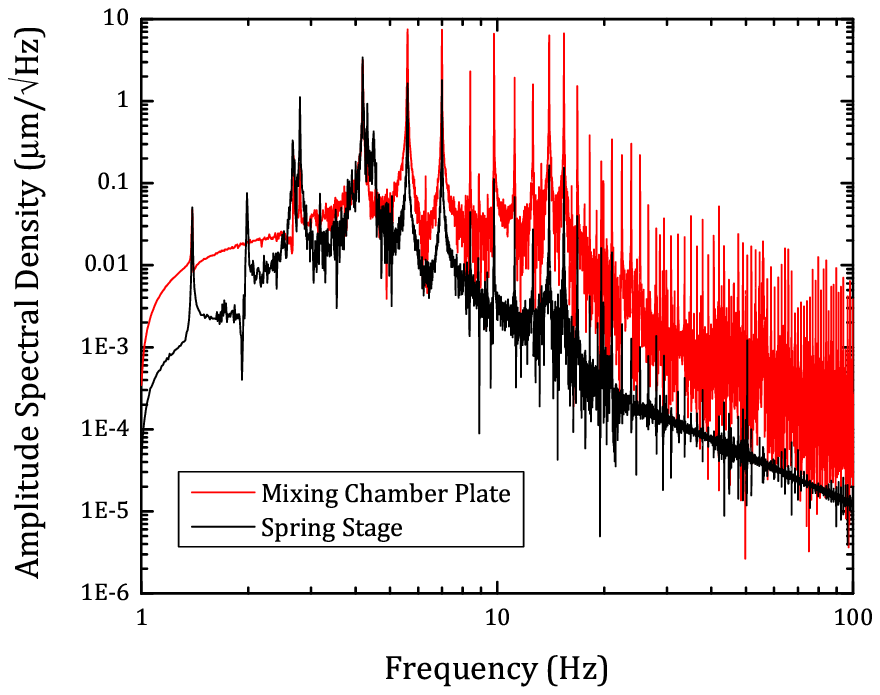}
\caption{Vibrations in the vertical direction on the mixing chamber plate and spring stage with the pulse tube on at room temperature. Above 10 Hz, there is a significant attenuation of the pulse tube harmonics at the spring stage.}
\label{fig:6}
\end{figure}
A comparison of the vibrations on the mixing chamber plate and spring stage highlights the function of the spring stage in the SGM design, and is shown in figure \ref{fig:6}. At 5 Hz and below, the spring stage gives almost no attenuation of the vibrations on the mixing chamber. The utility of the spring stage is at higher frequencies, especially above 10 Hz. The higher harmonics of the pulse tune vibration are significantly attenuated, up to a factor of $10^3$ at 100 Hz. The RMS vibration amplitude on the spring stage has been reduced to 0.58 $\mu$m, down from 1.9 $\mu$m at the mixing chamber plate. The main role of the spring stage however is to attenuate at frequencies at or above to the resonant frequency of the body of the scanner, which is well above 10 Hz. The series combination of a ``low-pass'' spring stage and a ``high-pass'' rigid scanner body is what ultimately allows for the relative tip-sample vibrations to be low.

\begin{figure}[fig7]
\centering
\includegraphics[width=0.8\textwidth]{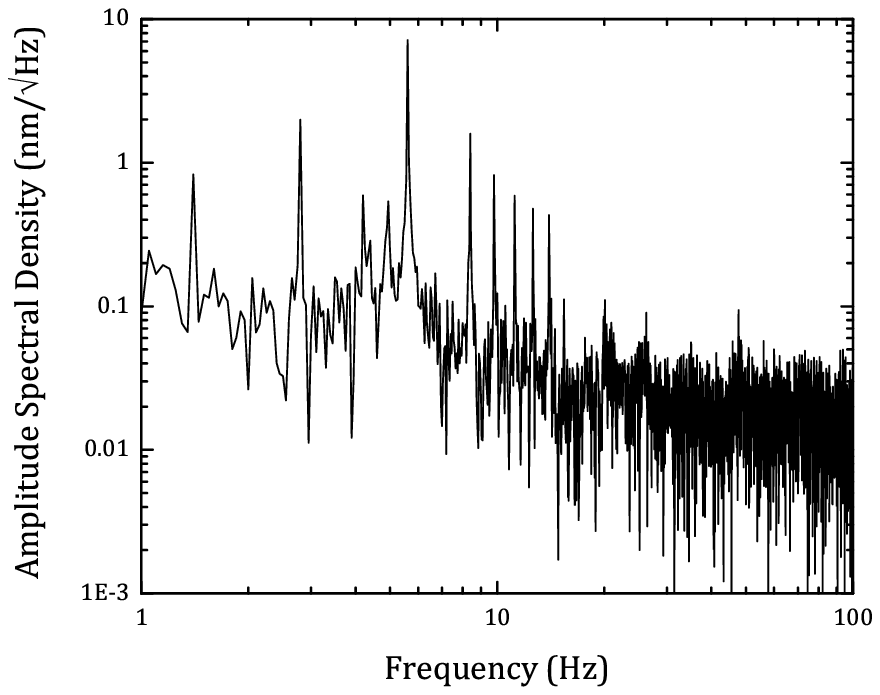}
\caption{Tip-sample vibrations while scanning at 15 mK. The integrated RMS vibration  between 1 Hz and 1 kHz at 15 mK is 2.1 nm.}
\label{fig:7}
\end{figure}
We now turn to a direct measurement of vibrations between the tip and the sample while scanning. This measurement is performed by operating the SGM in tapping mode, where the tip is in intermittent contact with the surface, at a large amplitude with no height feedback. For these measurements, the peak-to-peak oscillation height of the tip is 50 nm. By using a large amplitude excitation, vibrations that are much smaller than 50 nm will produce an approximately linear response in the tuning fork resonance. The response of the tuning fork is then measured with a phase locked loop, and any variation in the resonant frequency is a result of relative vibration between the tip and sample. The extension of the $z$ piezo is simultaneously oscillated at a known frequency and amplitude for calibration purposes, in this case 1.0 nm RMS at 25 Hz. The electrical response of the tuning fork can then be converted into a vibration amplitude by calibrating the power under the peak in the Fourier spectrum at 25 Hz. 

The result of the measurement is shown in figure \ref{fig:7}, with the calibration signal at 25 Hz removed. The integrated RMS vibration between 1 Hz and 1 kHz at 15 mK is 2.1 nm, a similar measurement performed at 3.3 K gives an RMS vibration amplitude of 0.8 nm. The increase in vibration at base temperature is a result of the mixture flowing in the dilution circuit. These vibrations are manageable for scanning gate measurements, where the tip is typically scanned at least 10 - 20 nm above the surface when gating. If a significantly lower vibration floor is needed for scanning, as would be required for STM measurements, the first modification to this design would be to mechanically decouple the pulse tube from the fridge as much as possible given the cooling power constraints of the system. For a system with a large superconducting magnet, this would require a design that allows for good thermal contact to the magnet while simultaneously providing a mechanically floppy contact to the scanner. Another improvement that could be made is the introduction of active vibration cancellation. Since the pulse tube vibrations are regular, one could implement a scheme that actively measures vibrations at the scanning stage and compensates for them through the motion of the scanning piezos.

\section{\label{sec:heatsink}WIRING AND FILTERING}
The wiring and low temperature filtering implemented in the SGM was chosen to optimize measurements of large sample impedances ($>$ 1 G$\Omega$) encountered in tunneling measurements\citep{Sciambi2010}. A number of different factors help determine the type of wiring and filtering that is implemented in a cryostat, and the presence of a pulse tube factors into these considerations for our SGM. One source of noise that is significant in cryogen-free systems is triboelectric or microphonic noise, which is the result of a physical vibration of the cable. For shielded cabling, vibrations can lead to friction between the inner/outer conductors and the dielectric, which can dump charge on the conductors and appear as excess current noise in a sensitive measurement. In a pulse tube based system, the hallmark of triboelectric noise is a peak in the noise spectrum at the pulse frequency since there is no purely electrical source of noise at this frequency, only mechanical. In fact, when making measurements of vibrations as described in section \ref{vibrations}, one must take care not to confuse triboelectric noise for tip-sample vibrations. These can be discriminated by comparing the noise spectrum with the tip in tapping contact and retracted from the surface.
\subsection{Sample Wiring}
The sample wiring can be divided into two main sections; wiring from room temperature to the mixing chamber, and wiring from the mixing chamber to the scanner. Between room temperature and the mixing chamber, there are 8 coaxial cables with a 36 AWG CuNi (C71500) inner conductor and 0.025'' outer diameter (OD) braided CuNi shield from Calmont Wire \& Cable. These coaxial cables have a graphite coating between the dielectric and outer shield to help reduce triboelectric noise. Care must be taken when using these cables to avoid shorts between the graphite and the inner conductor. There is an additional set of 24 PhBr twisted pair lines in a CuNi sheath from room temperature to the mixing chamber. These cables lead into a set of filtering and heat sinking stages on the mixing chamber, one of which is shown in figure \ref{fig:9}. After this stage, all cabling between the mixing chamber and scanner is Ag plated Cu coax with a braided shield. The sample is mounted on a chip carrier from Kyocera (PB-44567), and the chip socket is from Plastronics (P2032S-A-AU) with BeCu pins, as can be seen in figure \ref{fig:2}(b). Both the chip carrier and chip socket are non-magnetic, and are free of ferromagnetic underlayers beneath the gold contacts.

When considering voltage biased AC measurements, it is important to keep the pin-to-pin capacitance low to avoid a capacitive shunt across the device that has a comparable impedance to the device at the measurement frequency, which for us is typically below 100 Hz. For a 1 G$\Omega$ sample impedance and 100 Hz measurement frequency, this requires a pin-to-pin capacitance much smaller than $\left[2\pi(100\mbox{ Hz})(1\mbox{ G}\Omega)\right]^{-1} \approx 1.6\mbox{ pF}$. This requires the use of individually shielded cables, which is why we chose to have eight lines be coaxial from room temperature to the sample. The typical capacitance between two sample pins connected to the coaxial cabling is about 1 fF, limited only by the geometry of the chip carrier and chip socket. It is also important that these coaxes have low triboelectric noise, as this will appear as current noise in the conductance measurement. The twisted pair wiring is used for applying voltages to gates, where low pin-to-pin capacitance is less crucial. 

The Attocube coarse positioning stages require special consideration because there is a limit on the DC resistance of the cabling for proper stepper operation. At low temperature, the piezos in the steppers typically have about 150 nF of capacitance. The slip-stick motion relies on applying a fast voltage pulse to the piezo to rapidly expand or contract the piezo to overcome static friction. A total line resistance of less than 5 $\Omega$ is needed for operation to avoid a significant RC filtering of the pulse, but using copper cable would present a large heat load to the mixing chamber and is therefore avoided. To maintain low DC resistance while still providing a small heat load to the mixing chamber, superconducting coaxial cables with a NbTi inner conductor and CuNi outer conductor (Coax Co., Japan, SC-033/50-NbTi-CN) are used between the still and mixing chamber. From room temperature to the still, semi-rigid coaxial cables with a Ag plated BeCu inner conductor and CuNi outer conductor are used, which have a resistance of about 1 $\Omega$ each. A separate coaxial cable is used as the stepper ground, as tying the stepper ground directly to the fridge ground resulted in a cross-coupling between the slip-stick drive and the sample.
\subsection{RF Filtering and Heat Sinking}\label{rf}
\begin{figure}[fig9]
\centering
\includegraphics[width=0.8\textwidth]{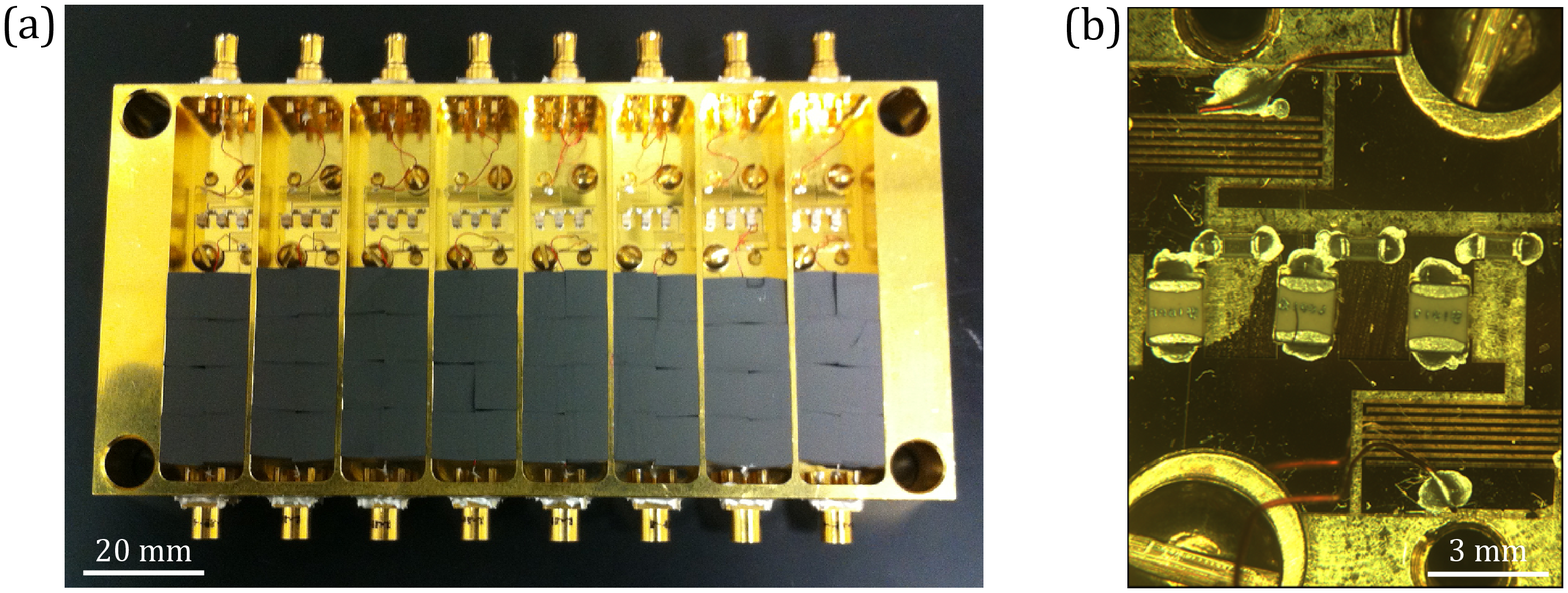}
\caption{Low temperature RF filtering and heat sinking stage. (a) The lower half of the stage contains Eccosorb foam for filtering above 1 GHz. Passive RC filters mounted on sapphire provide heat sinking in the upper half of the stage. (b) Close-up image of the RC filters mounted on sapphire. The sapphire is anchored to the box with two brass screws that also provide ground contact for the RC filters.}
\label{fig:9}
\end{figure}

To effectively thermalize the coaxial cables to the mixing chamber, the inner conductor should be broken out of the coax and heat sunk separately from the outer shield. The stage designed to do this is shown in figure \ref{fig:9}(a), which filters eight lines. This stage provides both heat sinking of the inner conductor and electrical filtering above 1 MHz. Seven of these stages are mounted in the system to filter the 32 sample lines and all lines associated with the Attocube coarse steppers and piezo scanner. Each stage is machined out of OFHC copper, and MCX connectors are used for the inputs and outputs.

\begin{figure}[fig10]
\centering
\includegraphics[width=0.8\textwidth]{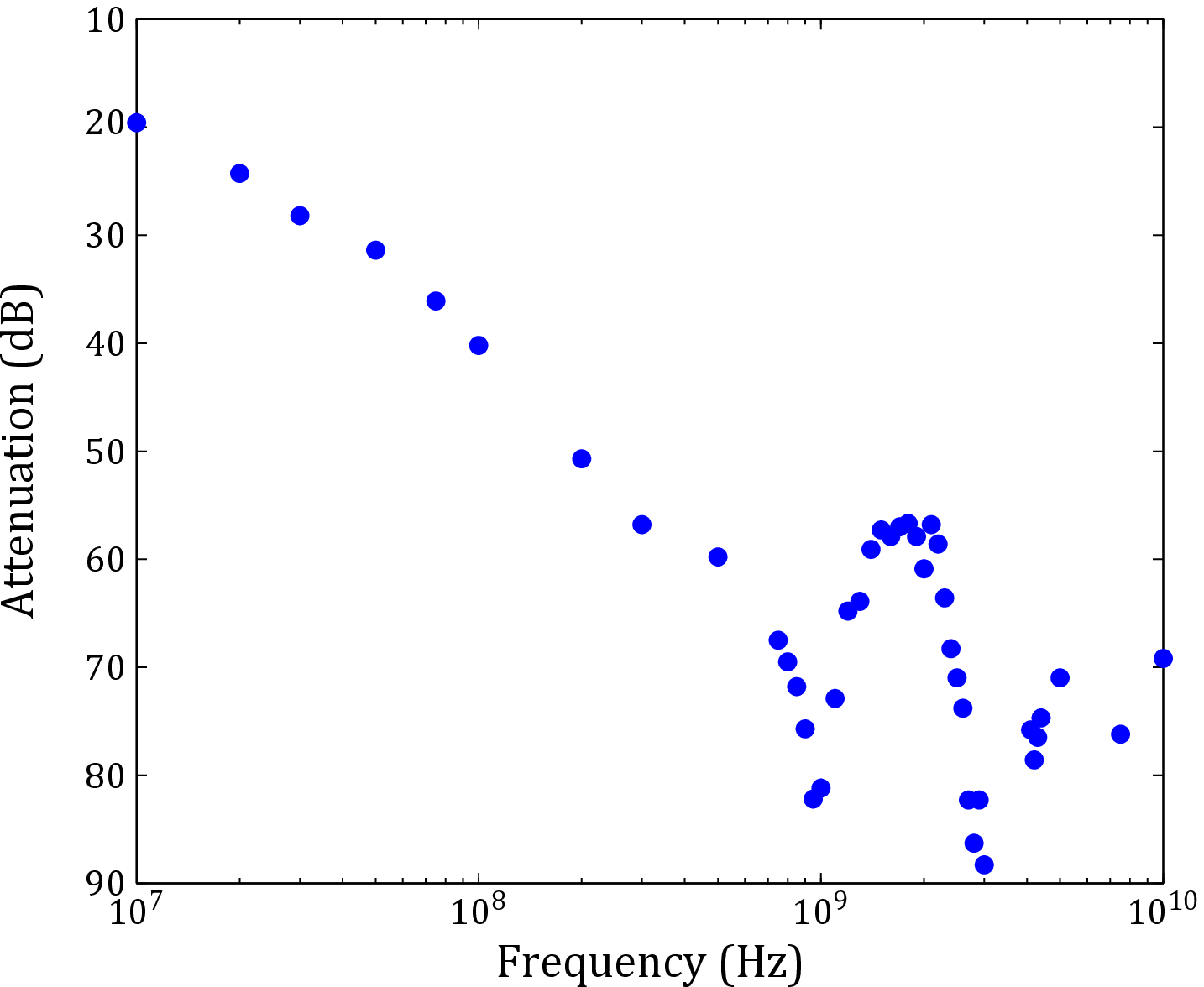}
\caption{Attenuation of the RF filtering stage versus frequency.}
\label{fig:10}
\end{figure}

There are two main sections to each filtering stage. The input is along the bottom of  figure \ref{fig:9}, and the lower half of the box provides filtering above 1 GHz using a dielectric that becomes lossy in this frequency range\citep{Santavicca2008,Slichter2009}. The dielectric used is Eccosorb foam (MFS-117), which has been shown to be a suitable replacement for traditional copper powder filters\citep{Santavicca2008} at magnetic fields below 0.1 T. It has also been shown to thermalize well down to 25 mK with comparable filtering performance at room temperature and low temperature\citep{Slichter2009}. The upper half of the box contains a 0.5 mm thick sapphire substrate on which a 3 stage RC filter and meander traces are patterned, as shown in figure \ref{fig:9}(b). The metal traces on the sapphire are Ti/Au (5 nm/200 nm) which was annealed at 450 C for 60 seconds after deposition. Meander traces are included before and after the RC filters to help heat sink the trace. The RC filters are each 200 $\Omega$ with 100 pF/10 pF/1 pF capacitors from ATC. The inline resistance of the RC stages along with the meander traces is approximately 700 $\Omega$ at room temperature, which changes by less than 5\% at base temperature. The RC filters are omitted for the Attocube coarse positioning stages to maintain a low DC resistance, however sapphire is still used for heat sinking these lines. The measured attenuation of the completed stage is shown in figure \ref{fig:10}.

To maintain a strong thermal connection between the sample and the heat sinking stages on the mixing chamber, no solder is used in the design of the heat sinking stages or in the path from the heat sinks to the sample. All electrical connections are made either by crimping, press fitting, or with silver epoxy (Epo-Tek H20E) in this region. Lead-tin solders will superconduct at these temperatures, which will act as effective breaks in the thermal anchoring unless an external magnetic field is applied to drive the solder normal.

\subsection{Electron Temperature Measurement}
\begin{figure}[fig8]
\centering
\includegraphics[width=0.8\textwidth]{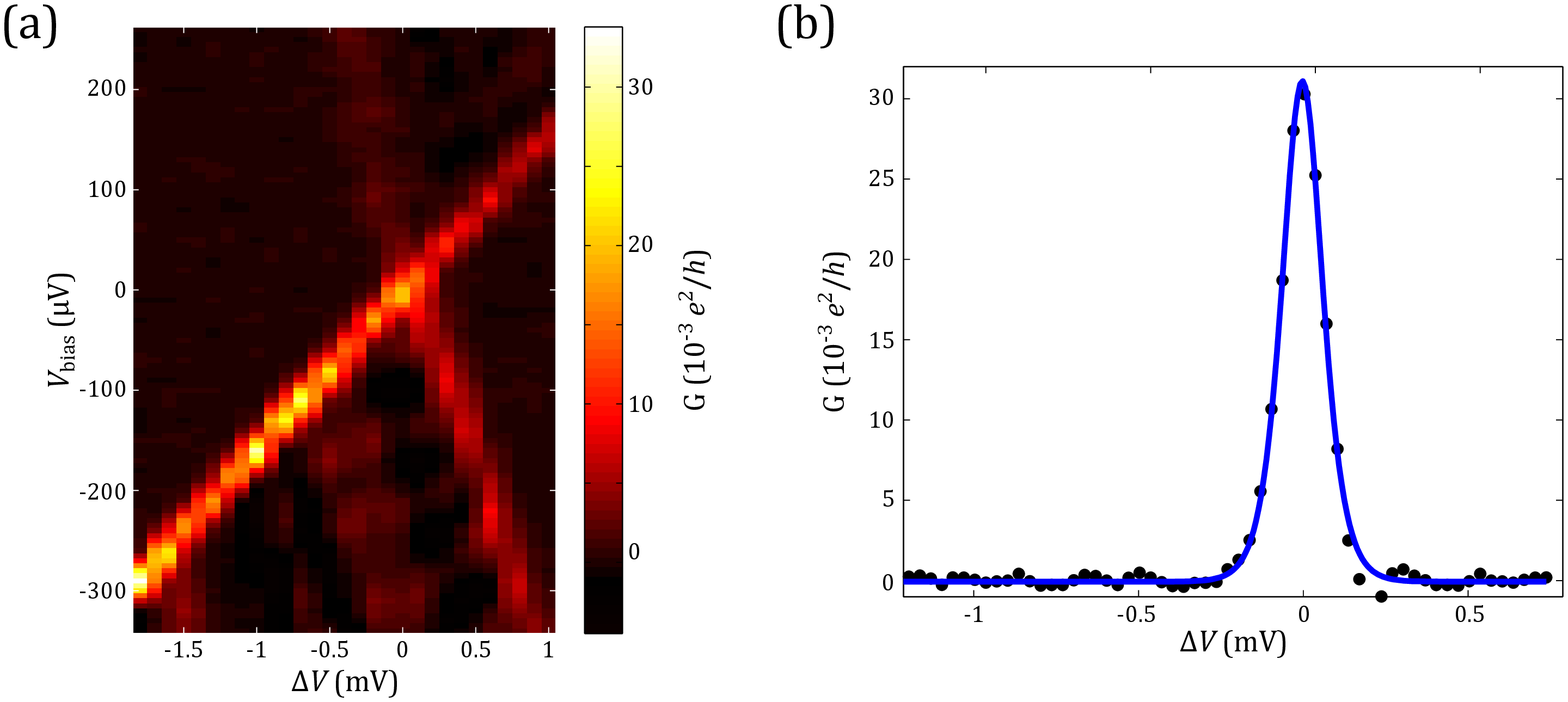}
\caption{Coulomb blockade thermometry on a quantum dot. (a) Differential conductance $G$ through the dot as a function of plunger gate voltage $\Delta V$ and source-drain bias $V_{bias}$. A fit to the slope of the Coulomb diamond gives $\alpha = 0.097$. (b) Differential conductance versus plunger gate voltage at zero source-drain bias. A fit to equation \ref{cosh} gives an electron temperature of $T_e = 45$ mK.}
\label{fig:8}
\end{figure}

The design of the low temperature filtering stages is geared toward achieving a low electron temperature in the sample. One method to directly measure electron temperature is to use a quantum dot and measure the electron temperature of the source and drain leads via Coulomb blockade thermometry. The width of a peak in differential conductance through the dot in the Coulomb blockade regime is temperature limited when the tunnel coupling $\Gamma$ of the dot to the source and drain leads is small ($\hbar \Gamma \ll k T$) and the level spacing of the dot is large ($kT\ll \Delta E$). The functional form of the differential conductance $G$ as a function of a plunger gate voltage $V_g$ is given by\citep{Beenakker1991}
\begin{eqnarray}\label{cosh}
G(V_g) &=& G_0\cosh^{-2}\left(\frac{e\alpha\Delta V}{2 kT}\right),
\end{eqnarray}
where $\Delta V = V_g - V_g^0$ with the peak centered at $V_g^0$, and $\alpha$ reflects the capacitive coupling between the plunger gate and the dot, which can be measured from the shape of a Coulomb diamond. The result of this measurement is shown in figure \ref{fig:8}.
The lowest electron temperature we observed was 45 mK. It should be noted that this measurement only provides an effective upper bound on the temperature, as other extrinsic noise sources could act to broaden the peak. In particular, triboelectric noise on the wires connected to the gates of the dot can act to broaden the energy level of the dot. However, the presence of coaxial cabling with a braided shield between the mixing chamber and the scanner may compromise the RF filtering to a certain degree. A braided shield was chosen to minimize vibrations transmitted by the wiring, as semi-rigid cabling would defeat the use of a spring supported stage for vibration isolation.

\section{Conclusions}
We have presented details of our design of a scanning gate microscope housed in a cryogen-free dilution refrigerator. The vibrations from the pulse tube are attenuated by our SGM design to 0.8 nm RMS at 3 K and 2.1 nm RMS at 15 mK, small enough for scanning gate measurements on mesoscopic electron systems. Further improvements need to be made to the design to achieve vibration levels suitable for STM measurements, which could possibly be achieved by mechanically decoupling the pulse tube from the cryostat. However, this would yield a reduction in cooling power that is not manageable with the cooling requirements for the vector magnet in our system. We hope that our demonstration of scanning in a cryogen-free system helps others who are considering or attempting to build similar systems, especially given the current trend towards dry systems in cryogenics.

\begin{acknowledgments}
This work is supported by DOE-BES, DMS\&E at SLAC (DE-AC02-76SF00515), with the original concept developed under the Center for Probing the Nanoscale (NSF NSEC Grant No. 0425897) and a Mel Schwartz Fellowship from the Stanford Physics Department. M.P. acknowledges support from the Hertz Foundation, NSF, and Stanford. D.G.-G. recognizes support from the David and Lucile Packard Foundation. We thank Z. Kermish for helpful discussions regarding the linear motor driver.
\end{acknowledgments}

\nocite{*}
\bibliography{rsibib}

\end{document}